\documentclass[american,english,12pt,ruled]{article}
\usepackage[T1]{fontenc}
\usepackage[utf8]{inputenc}
\usepackage{algorithm2e}
\usepackage{amsmath}
\usepackage[authoryear]{natbib}

\makeatletter
\@ifundefined{date}{}{\date{}}
\RequirePackage{graphicx,color,ae,fancyvrb}
\RequirePackage[T1]{fontenc}

\usepackage{amsthm,amsmath}
\usepackage{mathrsfs,amsfonts}
\usepackage{mathtools}
\usepackage{booktabs}
\usepackage[margin=1in]{geometry}
\usepackage{microtype}
\usepackage{setspace}
\usepackage{hyperref}
\definecolor{Red}{rgb}{0.5,0,0}
\definecolor{Blue}{rgb}{0,0,0.5}

\hypersetup{%
colorlinks = {true},
linktocpage = {true},
plainpages = {false},
linkcolor = {Blue},
citecolor = {Blue},
urlcolor = {Red},
}

\makeatother

\usepackage{babel}
\begin{document}
\selectlanguage{american}%
\global\long\def\ev{\mathsf{E}}%

\global\long\def\var{\mathsf{var}}%

\global\long\def\real{\mathbf{R}}%

\global\long\def\T{{\scriptscriptstyle \text{T}}}%

\global\long\def\pr{\mathsf{Pr}}%

\selectlanguage{english}%
\global\long\def\bs#1{\boldsymbol{#1}}%

\global\long\def\argmin{\operatorname*{arg\,min}}%

\global\long\def\argmax{\operatorname*{arg\,max}}%

\title{Dispersion Modeling in Zero-inflated Tweedie Models with Applications
to Insurance Claim Data Analysis}
\author{Yuwen Gu\\
Department of Statistics{\small}\\
University of Connecticut}
\maketitle
\begin{abstract}
The Tweedie generalized linear models are commonly applied in the
insurance industry to analyze semicontinuous claim data. For better
prediction of the aggregated claim size, the mean and dispersion of
the Tweedie model are often estimated together using the double generalized
linear models. In some actuarial applications, it is common to observe
an excessive percentage of zeros, which often results in a decline
in the performance of the Tweedie model. The zero-inflated Tweedie
model has been recently considered in the literature, which draws
inspiration from the zero-inflated Poisson model. In this article,
we consider the problem of dispersion modeling of the Tweedie state
in the zero-inflated Tweedie model, in addition to the mean modeling.
We also model the probability of the zero state based on the generalized
expectation-maximization algorithm. To potentially incorporate nonlinear
and interaction effects of the covariates, we estimate the mean, dispersion,
and zero-state probability using decision-tree-based gradient boosting.
We conduct extensive numerical studies to demonstrate the improved
performance of our method over existing ones.\bigskip{}

\end{abstract}
\doublespacing

\section{Introduction}

Semicontinuous data arise frequently from scientific, social and economic
studies. These data exhibit a unique pattern: besides a large portion
of exact zero values, they typically have a right-skewed shape for
the positive continuous values. For example, it is common in actuarial
practices, such as in auto, property and casualty insurance, to observe
semicontinuous claim sizes, where the zeros often correspond to policy
contracts under which no claims are filed. While for policies under
which claims do exist, the aggregated claim sizes may have a highly
right-skewed continuous distribution.

Accurate modeling of such semicontinuous data is often of vital importance
to the insurance industry. For example, it is crucial that an insurance
company set accurate premiums for its policyholders to avoid adverse
selection \citep{dionne01_testin_eviden_adver_selec_autom_insur_market}.
Indeed, a fair premium needs to be charged based on the potential
claim size of each policy. If an auto insurance company charges city
drivers much more than suburban drivers, then city drivers will turn
to competitors of the company so that the policies for its suburban
drivers will be underpriced. Thus, the insurance company is left with
risky policies and loses profitable ones, resulting in financial loss.
Therefore, to set an appropriate premium, the insurance company needs
to accurately model the semicontinuous claim size of each policy.
However, special care must be taken when dealing with such semicontinuous
data, since a simple transformation to normality would not work well
due to the probability mass at zero.

In the realm of semicontinuous data modeling, four primary approaches
are commonly employed. The Tobit model \citep{tobin58estimation}
belongs to a class of regression models where the observed range of
the dependent variable is censored or truncated \citep{Hayashi2000econ}.
The Hurdle model \citep{gragg1971some} is similar to the Tobit model.
In the hurdle model, separate equations are used for bounded and unbounded
outcomes, while the Tobit model uses a single equation for both. The
hurdle model assumes that the unbounded outcomes occur when a hurdle
is cleared, while bounded outcomes result when the hurdle is not cleared.
The two-part model \citep{manning1981two}, comprising two distinct
parts, is also commonly used. In the first part, a random variable
is considered to determine whether an observation is zero or nonzero,
while the second part involves another random variable describing
the magnitude of the positive observation. Lastly, the Tweedie model
\citep{tweedie1984index}, derived from a compound mixture of Poisson
and gamma, is probably one of the most commonly used methods in the
insurance industry. The model assumes Poisson arrival of events and
gamma distributed sizes for individual events. This interpretation
is of particular interest in actuarial applications, where the Poisson
part describes claim frequency and the gamma part models claim severity.

The Tweedie generalized linear models (GLMs) are extensively used
in actuarial studies \citep{joergensen94_fittin_tweed_compoun_poiss_model,qian16_tweed_compoun_poiss_model_with}
for Tweedie distribution's intimate connection to the exponential
dispersion models \citep{jorgensen1997_theory_disper_models,dunn05_series_evaluat_tweed_expon_disper_model_densit},
where a logarithmic link function is often employed to connect the
mean aggregated claim size to a linear function of available covariates.
Two possible improvements can often be made to more accurately model
the mean of the Tweedie model. Firstly, the linear structure for the
transformed mean may become too restrictive for some applications
where interpretability is not of the primary concern. To that end,
the Tweedie model can be extended to incorporate nonlinearities and/or
interactions using splines \citep{zhang2013likelihood}, additive
models \citep{chen2023generalized}, and boosting \citep{yang2018insurance}.
Secondly, it is well known that accurate estimation of the dispersion
can often help improve the estimation and inference of the mean \citep{smyth99_adjus_likel_method_model_disper}.
Moreover, to better understand the underlying frequency and severity
sub-models within the Tweedie model, it is generally necessary to
model the dispersion of the claim sizes as well as their mean. To
that end, the double generalized linear models can be used and it
is believed that dispersion modeling in addition to the mean modeling
generally helps improve the accuracy of the Tweedie model \citep{smyth99_adjus_likel_method_model_disper,smyth02_fittin_tweed_poiss}.

Despite the popularity of the Tweedie GLMs and their various extensions,
it is not uncommon to see the Tweedie models perform badly when the
percentage of zeros is extremely high (e.g., greater than 90\%) in
the data. The excess zeros can inflate the dispersion estimation of
the Tweedie model which in turn deteriorates the accuracy of the mean
estimation. Motivated by the zero-inflated Poisson model \citep{lambert92_zero_inflat_poiss_regres_with},
the zero-inflated Tweedie model is proposed by \citet{zhou20_tweed_gradien_boost_extrem_unbal}
to handle the excessive proportions of zeros. Specifically, to fit
the zero-inflated Tweedie model, \citet{zhou20_tweed_gradien_boost_extrem_unbal}
assume a varying mean of the Tweedie state but a non-varying dispersion
and non-varying probability of the zero-state. In this article, we
improve the estimation approach of \citet{zhou20_tweed_gradien_boost_extrem_unbal}
by considering varying zero-state probability and varying dispersion
in addition to the varying mean. Using the generalized expectation-maximization
\citep[EM,][]{dempster77_maxim_likel_from_incom_data_via_algor,borman2004expectation}
algorithm, we iteratively estimate the mean, dispersion, and zero-state
probability non-parametrically using decision-tree-based gradient
boosting. As a result, our proposed method naturally handles various
types of covariates (numeric or categorical), and missing values and
outliers therein by virtue of the decision trees.

The remainder of this article is organized as follows. In Section
\ref{sec:tweedie}, we review the Tweedie model and introduce the
problem of dispersion modeling in the Tweedie model under the double
generalized linear models framework. We introduce the zero-inflated
Tweedie model and consider nonparametric dispersion modeling therein
in Section \ref{sec:zit}, where in addition, we also model its mean
and zero-state probability non-parametrically. We conduct extensive
simulation experiments as well as analyze a real-world auto insurance
dataset in Section \ref{sec:numerical-studies} to demonstrate the
competitive performance of our method. Finally, we make some conclusive
remarks in Section \ref{sec:conclusions}.

\section{The Tweedie model\protect\label{sec:tweedie}}

Let $N_{i}$ be the number of claims observed under the $i$th policy,
and $Z_{i}$ be the total claim size for that policy. Suppose that
the exposure (typically measured in policy years) of the $i$th policy
is $w_{i}$, and let $Y_{i}=Z_{i}/w_{i}$ be the observed total claim
size per unit exposure. Assume that $N_{i}$ follows a Poisson distribution
with mean $\lambda_{i}w_{i}$, so $N_{i}$ and $Y_{i}$ are zero with
probability $\exp(-\lambda_{i}w_{i})$. Assume also that individual
claims under policy $i$ arrive independently and the size of each
claim is gamma distributed with mean $\tau_{i}$ and shape parameter
$\alpha$. Thus, conditional on $N_{i}>0$, $Y_{i}$ is continuous
and positive, and has the gamma distribution with mean $N_{i}\tau_{i}/w_{i}$.
It is well known that $Y_{i}$ follows a Tweedie distribution \citep{tweedie1984index,jorgensen1997_theory_disper_models}
with mean parameter $\mu_{i}=\ev(Y_{i})=\lambda_{i}\tau_{i}$ , power
parameter $\zeta=(\alpha+2)/(\alpha+1)$, and dispersion parameter
\[
\phi_{i}=w_{i}\frac{\var(Y_{i})}{\mu_{i}^{\zeta}}=\frac{\lambda_{i}^{1-\zeta}\tau_{i}^{2-\zeta}}{2-\zeta}.
\]
The distribution of $Y_{i}$, denoted by $\text{Tweedie}(\mu_{i},\phi_{i}/w_{i},\zeta)$
in the sequel, has the following density form
\begin{equation}
d_{\text{TD}}(y;\mu_{i},\phi_{i}/w_{i},\zeta)=\exp\left\{ \frac{w_{i}}{\phi_{i}}\left(\frac{y\mu_{i}^{1-\zeta}}{1-\zeta}-\frac{\mu_{i}^{2-\zeta}}{2-\zeta}\right)+c\left(y,\frac{\phi_{i}}{w_{i}},\zeta\right)\right\} ,\label{eq:tweedie-density}
\end{equation}
where
\[
c(y,\phi,\zeta)=\begin{cases}
0, & \text{if }y=0,\\
\log\left[\frac{1}{y}\sum_{j=1}^{\infty}\frac{y^{j\alpha}}{\phi^{j(1+\alpha)}(2-\zeta)^{j}(\zeta-1)^{j\alpha}j!\Gamma(j\alpha)}\right], & \text{if }y>0,
\end{cases}
\]
and $\Gamma(\cdot)$ is the gamma function. 

Given $\zeta\in(1,2)$, the Tweedie distribution belongs to the family
of exponential dispersion models \citep{jorgensen1997_theory_disper_models}.
In light of (\ref{eq:tweedie-density}), the distribution has a positive
probability mass $\pr(Y_{i}=0)=\exp(-\lambda_{i}w_{i})=\exp\{-w_{i}\mu_{i}^{2-\zeta}/[\phi_{i}(2-\zeta)]\}$
at zero and a continuous density on the positive reals. Th compound
Poisson-gamma mixture makes the Tweedie model an attractive tool for
analyzing insurance claim data, where frequency and severity of the
claims are simultaneously captured. In particularly, the Poisson part
takes care of the exact zero values from policies where no claims
are filed, and the gamma part takes care of the often observed right-skewed
distribution of the claim sizes. Compared to other models for semicontinuous
data, such as the two-part model and the Tobit model \citep{liu19_statis_analy_zero_inflat_nonneg_contin_data},
the Tweedie model bears a more natural interpretation.

\subsection{Dispersion modeling}

Note that $\zeta\in(1,2)$ since $\alpha>0$. As \citet{smyth02_fittin_tweed_poiss}
point out, any factor that affects either the underlying frequency
$\lambda_{i}$ of claims or their average size (severity) $\tau_{i}$
would likely affect both the mean $\mu_{i}=\lambda_{i}\tau_{i}$ and
the dispersion $\phi_{i}=\lambda_{i}^{1-\zeta}\tau_{i}^{2-\zeta}/(2-\zeta)$.
Any factor that affects the mean but not the dispersion must affect
$\lambda_{i}$ and $\tau_{i}$ in such a way that $\lambda_{i}^{1-\zeta}\tau_{i}^{2-\zeta}$
remains constant, which is quite unlikely. Therefore, it is reasonable
to allow both $\mu_{i}$ and $\phi_{i}$ to depend on the values of
the covariates. In this article, we focus on the dispersion modeling
in addition to the regular mean modeling, as functions of the covariates.
On the one hand, it is well known that efficient estimation of the
mean parameters in a regression model depends on accurate modeling
of the dispersion \citep{smyth99_adjus_likel_method_model_disper}.
The loss of efficiency in using a constant dispersion when the dispersion
is varying could be substantial, which may lead to inaccurate inference,
such as unreliable standard errors and confidence intervals, on the
mean parameters. On the other hand, the dispersion may be of direct
interest on its own, such as in risk management, or in understanding
both the underlying frequency and severity of the claims. Hence, given
the vector of covariates $\bs x_{i}$ for the $i$th policy, we assume
\[
g(\mu_{i})=F_{\mu}(\bs x_{i})\enskip\text{and}\enskip h(\phi_{i})=F_{\phi}(\bs x_{i}),
\]
where $g$ and $h$ are known monotonic link functions, and $F_{\mu}$
and $F_{\phi}$ are unknown functions to be estimated non-parametrically
from the data.

The normalizing function $c(y,\phi_{i}/w_{i},\zeta)$ in (\ref{eq:tweedie-density})
can be a complicated function of $\phi_{i}$, which makes simultaneous
estimation of $F_{\mu}$ and $F_{\phi}$ difficult, even when $F_{\mu}$
and $F_{\phi}$ are both assumed to be of simple forms, such as linear
functions, of the covariates. In practice, we can consider the saddle-point
approximation $c(y,\phi,\zeta)\approx-\log(2\pi\phi y^{\zeta})/2$,
which holds well when $\phi$ is small \citep{jorgensen1997_theory_disper_models}.
The accuracy of the saddle-point approximation is carefully studied
in \citet{smyth99_adjus_likel_method_model_disper}. The saddle-point
approximation also has the interpretation of being connected to an
extended quasi-likelihood \citep{nelder87_exten_quasi_likel_funct},
which may be less subject to model misspecification than the likelihood
(\ref{eq:tweedie-density}). The approximation gives rise to the zigzag
iterations \citep{smyth96_partit_algor_maxim_likel_other}, which
update $F_{\mu}$ and $F_{\phi}$ in an alternating fashion. In particular,
in the double generalized linear models framework \citep{nelder87_exten_quasi_likel_funct,smyth99_adjus_likel_method_model_disper,smyth02_fittin_tweed_poiss},
one assumes $F_{\mu}(\bs x)=\bs x^{\T}\bs{\beta}$ and $F_{\phi}(\bs x)=\bs x^{\T}\bs{\lambda}$.
The mean parameters $\bs{\beta}$ and the dispersion parameters $\bs{\lambda}$
are then updated alternately using the zigzag iterations. Specifically,
one can show that given $\bs{\lambda}$, $\bs{\beta}$ can be obtained
from solving a generalized linear model based on the deviance function
of the Tweedie model, and given $\bs{\beta}$, $\bs{\lambda}$ can
be obtained from solving a gamma generalized linear model. The linearity
assumption, although very commonly made in many applications, may
fail to capture more complicated nonlinear and interaction effects.
Therefore, in this paper, we use decision-tree-based gradient boosting
to more flexibly estimate $F_{\mu}$ and $F_{\phi}$.

\section{Zero-inflated Tweedie model\protect\label{sec:zit}}

Despite Tweedie model's popular use in insurance claim data analysis
and beyond, its deteriorated performance has been observed when the
proportion of zeros is excessive \citep{zhou20_tweed_gradien_boost_extrem_unbal}.
To that end, we follow \citet{zhou20_tweed_gradien_boost_extrem_unbal}
and consider the zero-inflated Tweedie model
\begin{equation}
Y_{i}=\begin{cases}
0, & \text{with probability }\pi_{i},\\
T_{i}, & \text{with probability }1-\pi_{i},
\end{cases}\label{eq:zit-model}
\end{equation}
where $T_{i}\sim\text{Tweedie}(\mu_{i},\phi_{i}/w_{i},\zeta)$. The
added probability $\pi_{i}$, representing the chance of policy $i$
falling into the perfect, zero state, is used to accommodate the zero
inflation so that
\begin{align*}
 & \pr(Y_{i}=0)=\pi_{i}+(1-\pi_{i})\exp\left\{ -\frac{w_{i}\mu_{i}^{2-\zeta}}{\phi_{i}(2-\zeta)}\right\} \\
 & =\exp\left\{ -\frac{w_{i}\mu_{i}^{2-\zeta}}{\phi_{i}(2-\zeta)}\right\} +\pi_{i}\left[1-\exp\left\{ -\frac{w_{i}\mu_{i}^{2-\zeta}}{\phi_{i}(2-\zeta)}\right\} \right]>\pr(T_{i}=0).
\end{align*}
We remark that the zero-inflated Tweedie model bears a natural interpretation
of being a compound mixture of a zero-inflated Poisson \citep{lambert92_zero_infla_poisso_regres_applic_defect_manufa}
for frequency and gamma for severity. This interpretation makes the
zero-inflated Tweedie model our preferred approach to modeling aggregated
claim sizes with excess zeros, compared to alternative methods \citep[see, e.g.,][]{liu19_statis_analy_zero_inflat_nonneg_contin_data}
such as the Tobit and two-part models.

We assume a constant power parameter $\zeta$ across the policies,
which is common in the literature \citep{joergensen94_fittin_tweed_compoun_poiss_model,smyth02_fittin_tweed_poiss}.
However, for maximal model flexibility, we allow varying mean and
dispersion from policy to policy. The likelihood with policy $i$
is then
\[
f_{Y_{i}}(y;\pi_{i},\mu_{i},\phi_{i}/w_{i},\zeta)=\pi_{i}I(y=0)+(1-\pi_{i})d_{\text{TD}}(y;\mu_{i},\phi_{i}/w_{i},\zeta),
\]
where the expression of $d_{\text{TD}}$ is given in (\ref{eq:tweedie-density}).

Assume we are given a portfolio of $n$ independent policies $(y_{i},\bs x_{i},w_{i})$
for $i=1,\ldots,n$. For ease of notation, let $\bs y=(y_{1},\ldots,y_{n})^{\T}$,
$\bs X=(\bs x_{1},\ldots,\bs x_{n})^{\T}$ and $\bs w=(w_{1},\ldots,w_{n})^{\T}$.
Under the zero-inflated Tweedie model, the joint log-likelihood with
the observed data can be written as
\begin{equation}
\begin{aligned} & \ell(\theta,\zeta|\bs y,\bs X,\bs w)=\sum_{i=1}^{n}\log[p(y_{i}|\bs x_{i},w_{i})]=\sum_{i=1}^{n}\log f_{Y_{i}}(y_{i};\pi(\bs x_{i}),\mu(\bs x_{i}),\phi(\bs x_{i})/w_{i},\zeta)\\
 & =\sum_{i\colon y_{i}=0}\log\left(\pi(\bs x_{i})+(1-\pi(\bs x_{i}))\exp\left\{ -\frac{w_{i}[\mu(\bs x_{i})]^{2-\zeta}}{\phi(\bs x_{i})(2-\zeta)}\right\} \right)+\sum_{i\colon y_{i}>0}\log(1-\pi(\bs x_{i}))\\
 & \quad\,+\sum_{i\colon y_{i}>0}\left\{ \frac{w_{i}}{\phi(\bs x_{i})}\left(\frac{y_{i}[\mu(\bs x_{i})]^{1-\zeta}}{1-\zeta}-\frac{[\mu(\bs x_{i})]^{2-\zeta}}{2-\zeta}\right)+c\left(y_{i},\frac{\phi(\bs x_{i})}{w_{i}},\zeta\right)\right\} ,
\end{aligned}
\label{eq:zit-loglik}
\end{equation}
where $\theta=(F_{\mu},F_{\phi},F_{\pi})$ denotes the collection
of functional parameters, $\mu(\bs x_{i})=g^{-1}(F_{\mu}(\bs x_{i}))$,
$\phi(\bs x_{i})=h^{-1}(F_{\phi}(\bs x_{i}))$ and $\pi(\bs x_{i})=q^{-1}(F_{\pi}(\bs x_{i}))$,
and $g,h$ and $q$ are known monotonic link functions. Here, $F_{\mu},F_{\phi}$
and $F_{\pi}$ are unknown functions to be estimated and $\zeta$
is the scalar power parameter also to be estimated. In this article,
we choose $g$ and $h$ as the logarithmic function, and $q$ as the
logit function, $q(u)=\log[u/(1-u)]$. The unknown functions $F_{\mu},F_{\phi}$
and $F_{\pi}$ are estimated non-parametrically by decision-tree-based
gradient boosting, which is described in Section \ref{subsec:gbdt}.

We highlight a few key differences between our approach and the one
by \citet{zhou20_tweed_gradien_boost_extrem_unbal}. Firstly, as mentioned
in Section \ref{sec:tweedie}, this article will focus on dispersion
modeling of the Tweedie state in the zero-inflated model, to better
assist the estimation of the mean function, which in turn helps to
better predict the aggregated claim size. Secondly, we allow a varying
probability, $\pi_{i}$, of falling into the perfect zero state for
each policy $i$ and estimate it non-parametrically using gradient
boosted trees as well. Compared to \citet{zhou20_tweed_gradien_boost_extrem_unbal}
who assume $\pi_{i}$'s non-varying across all policies, this varying
zero-state probability assumption is more reasonable in practice for
modeling zero inflation. See, for example, the zero-inflated Poisson
model \citep{lambert92_zero_inflat_poiss_regres_with}, in which the
probability of falling into the zero state is also assumed to be dependent
on the covariates, albeit in a GLM setup. Lastly, we fit the gradient
boosted trees using the state-of-the-art package LightGBM \citep{ke2017lightgbm},
which is among the fastest gradient boosting implementations. Moreover,
we employ a sensible initialization of the EM algorithm based on the
zero-truncated Tweedie model (see Section \ref{subsec:em-Initialization})
that considerably reduces the number of iterations needed.

\subsection{Expectation-maximization algorithm\protect\label{subsec:em}}

Directly maximizing (\ref{eq:zit-loglik}) with respect to $\theta$
is challenging due to the sum of exponential terms in the logarithmic
part. To that end, we consider the EM algorithm \citep{dempster77_maxim_likel_from_incom_data_via_algor}.
Suppose we knew which zeros came from the perfect, zero state and
which zeros came from the Tweedie, i.e., suppose we could observe
$L_{i}=1$ when $Y_{i}$ is from the perfect state, and $L_{i}=0$
when $Y_{i}$ is from the Tweedie state. Then, we would have $(L_{i}|\bs x_{i})\sim\text{Bernoulli}(\pi(\bs x_{i}))$
and the log-likelihood with the ``complete'' data $(y_{i},\bs x_{i},w_{i},l_{i}),i=1,\ldots,n$
would be
\begin{equation}
\begin{aligned} & \ell_{C}(\theta,\zeta|\bs y,\bs X,\bs w,\bs l)=\sum_{i=1}^{n}\log[p(y_{i},l_{i}|\bs x_{i},w_{i})]\\
 & =\sum_{i=1}^{n}\biggl[(1-l_{i})\left\{ \frac{w_{i}}{\phi(\bs x_{i})}\left(\frac{y_{i}[\mu(\bs x_{i})]^{1-\zeta}}{1-\zeta}-\frac{[\mu(\bs x_{i})]^{2-\zeta}}{2-\zeta}\right)+c\left(y_{i},\frac{\phi(\bs x_{i})}{w_{i}},\zeta\right)\right\} \\
 & \qquad\quad\,+l_{i}\log(\pi(\bs x_{i}))+(1-l_{i})\log(1-\pi(\bs x_{i}))\biggr],
\end{aligned}
\label{eq:cmp-loglik}
\end{equation}
where $\bs l=(l_{1},\ldots,l_{n})^{\T}$. This log-likelihood is easier
to maximize, because $\pi(\cdot)$ is separated from $(\mu(\cdot),\phi(\cdot))$.
With fixed $\zeta$, we give the details of the EM algorithm for fitting
(\ref{eq:zit-loglik}) as follows. Suppose the EM algorithm is entering
iteration $(k+1)$, given update $\widehat{\theta}^{(k)}=(\widehat{F}_{\mu}^{(k)},\widehat{F}_{\phi}^{(k)},\widehat{F}_{\pi}^{(k)})$
from iteration $k$, where $k\geq0$.

\emph{E Step.}\quad{}We first calculate the posterior mean of $L_{i}$,
\begin{align*}
 & \Pi_{i}^{(k)}=\pr(l_{i}=1|y_{i},\bs x_{i},w_{i},\widehat{\theta}^{(k)})=\frac{p(y_{i}|l_{i}=1,\bs x_{i},w_{i,}\widehat{\theta}^{(k)})\pr(l_{i}=1|\bs x_{i},w_{i},\widehat{\theta}^{(k)})}{\sum_{l=0}^{1}p(y_{i}|l_{i}=l,\bs x_{i},w_{i},\widehat{\theta}^{(k)})\pr(l_{i}=l|\bs x_{i},w_{i},\widehat{\theta}^{(k)})}\\
 & =\frac{I(y_{i}=0)\widehat{\pi}^{(k)}(\bs x_{i})}{I(y_{i}=0)\widehat{\pi}^{(k)}(\bs x_{i})+[1-\widehat{\pi}^{(k)}(\bs x_{i})]d_{\text{TD}}(y_{i};\widehat{\mu}^{(k)}(\bs x_{i}),\widehat{\phi}^{(k)}(\bs x_{i})/w_{i},\zeta)}\\
 & =\begin{cases}
\frac{\widehat{\pi}^{(k)}(\bs x_{i})}{\widehat{\pi}^{(k)}(\bs x_{i})+[1-\widehat{\pi}^{(k)}(\bs x_{i})]\exp\left(-\frac{w_{i}[\widehat{\mu}^{(k)}(\bs x_{i})]^{2-\zeta}}{\widehat{\phi}^{(k)}(\bs x_{i})(2-\zeta)}\right)}, & \text{if }y_{i}=0,\\
0, & \text{if }y_{i}>0.
\end{cases}
\end{align*}
Therefore, the Q-function can be obtained from (\ref{eq:cmp-loglik})
as
\begin{equation}
\begin{aligned} & Q(\theta|\widehat{\theta}^{(k)})=\ev_{\bs l|\bs y,\bs X,\bs w,\widehat{\theta}^{(k)}}\left\{ \sum_{i=1}^{n}\log[p(y_{i},l_{i}|\bs x_{i},w_{i})]\right\} \\
 & =\sum_{i=1}^{n}\biggl[(1-\Pi_{i}^{(k)})\left\{ \frac{w_{i}}{\phi(\bs x_{i})}\left(\frac{y_{i}[\mu(\bs x_{i})]^{1-\zeta}}{1-\zeta}-\frac{[\mu(\bs x_{i})]^{2-\zeta}}{2-\zeta}\right)+c\left(y_{i},\frac{\phi(\bs x_{i})}{w_{i}},\zeta\right)\right\} \\
 & \qquad\quad\,+\Pi_{i}^{(k)}\log(\pi(\bs x_{i}))+(1-\Pi_{i}^{(k)})\log(1-\pi(\bs x_{i}))\biggr].
\end{aligned}
\label{eq:q-func}
\end{equation}

\emph{M Step for $F_{\pi}$.}\quad{}Find $\widehat{F}_{\pi}^{(k+1)}(\bs x)$
by maximizing
\begin{equation}
L_{C}(F_{\pi};\bs y,\bs X,\bs w,\bs{\Pi}^{(k)})=\sum_{i=1}^{n}\left[\Pi_{i}^{(k)}\log(\pi(\bs x_{i}))+(1-\Pi_{i}^{(k)})\log(1-\pi(\bs x_{i}))\right]\label{eq:em-pi-update}
\end{equation}
with respect to $F_{\pi}$, where $\pi(\bs x)=q^{-1}(F_{\pi}(\bs x))$
and $\bs{\Pi}^{(k)}=(\Pi_{1}^{(k)},\ldots,\Pi_{n}^{(k)})^{\T}$. In
this article, we take $q$ to be the logit link so that $\pi(\bs x)=[1+\exp(-F_{\pi}(\bs x))]^{-1}$.
As mentioned earlier, we will estimate $F_{\pi}$ non-parametrically
using gradient boosting, which is elaborated in Section \ref{subsec:gbdt}.

\emph{M Step for $F_{\mu}$ and $F_{\phi}$.}\quad{}Find $\widehat{F}_{\mu}^{(k+1)}(\bs x)$
and $\widehat{F}_{\phi}^{(k+1)}(\bs x)$ by maximizing
\begin{align*}
 & L_{C}(F_{\mu},F_{\phi};\bs y,\bs X,\bs w,\bs{\Pi}^{(k)})\\
 & =\sum_{i=1}^{n}(1-\Pi_{i}^{(k)})\left\{ \frac{w_{i}}{\phi(\bs x_{i})}\left(\frac{y_{i}[\mu(\bs x_{i})]^{1-\zeta}}{1-\zeta}-\frac{[\mu(\bs x_{i})]^{2-\zeta}}{2-\zeta}\right)+c\left(y_{i},\frac{\phi(\bs x_{i})}{w_{i}},\zeta\right)\right\} 
\end{align*}
with respect to $F_{\mu}$ and $F_{\phi}$, where $\mu(\bs x)=g^{-1}(F_{\mu}(\bs x))$
and $\phi(\bs x)=h^{-1}(F_{\phi}(\bs x))$. In this article, we take
$g$ and $h$ to be the logarithmic link so that
\[
\mu(\bs x)=\exp(F_{\mu}(\bs x))\enskip\text{and}\enskip\phi(\bs x)=\exp(F_{\phi}(\bs x)).
\]
Because $c(y_{i},\phi(\bs x_{i})/w_{i},\zeta)$ is complicated, simultaneous
estimation of $F_{\mu}$ and $F_{\phi}$ is difficult (see also Section
\ref{sec:tweedie}). We thus consider the following approximation
via extended quasi-likelihood \citep{nelder87_exten_quasi_likel_funct}
\begin{align*}
 & L_{C}^{+}(F_{\mu},F_{\phi};\bs y,\bs X,\bs w,\bs{\Pi}^{(k)})\\
 & =\sum_{i=1}^{n}(1-\Pi_{i}^{(k)})\left\{ -\frac{1}{2}\frac{w_{i}}{\phi(\bs x_{i})}D_{\zeta}(y_{i};\mu(\bs x_{i}))-\frac{1}{2}\log\left(2\pi\frac{\phi(\bs x_{i})}{w_{i}}y_{i}^{\zeta}\right)\right\} ,
\end{align*}
from which we can alternately update $F_{\mu}$ and $F_{\phi}$, where
\[
D_{\zeta}(y_{i};\mu(\bs x_{i}))=2\left[y_{i}\frac{y_{i}^{1-\zeta}-[\mu(\bs x_{i})]^{1-\zeta}}{1-\zeta}-\frac{y_{i}^{2-\zeta}-[\mu(\bs x_{i})]^{2-\zeta}}{2-\zeta}\right],i=1,\ldots,n
\]
are the unit deviances. The extended quasi-likelihood relies only
on moment assumptions, which may be less prone to model misspecification.
Specifically, given $F_{\phi}=\widehat{F}_{\phi}$ or $\phi=\widehat{\phi}$,
we update $F_{\mu}$ by maximizing
\begin{equation}
L_{C}^{+}(F_{\mu};\bs y,\bs X,\bs w,\bs{\Pi}^{(k)},\widehat{F}_{\phi})=-\frac{1}{2}\sum_{i=1}^{n}\frac{(1-\Pi_{i}^{(k)})w_{i}}{\widehat{\phi}(\bs x_{i})}D_{\zeta}(y_{i};\mu(\bs x_{i})),\label{eq:em-mu-update}
\end{equation}
which is based on a weighted Tweedie deviance with observational weights
$(1-\Pi_{i}^{(k)})w_{i}/\widehat{\phi}(\bs x_{i})$. Given $F_{\mu}=\widehat{F}_{\mu}$
or $\mu=\widehat{\mu}$, we update $F_{\phi}$ by maximizing
\begin{equation}
L_{C}^{+}(F_{\phi};\bs y,\bs X,\bs w,\bs{\Pi}^{(k)},\widehat{F}_{\mu})=\sum_{i=1}^{n}(1-\Pi_{i}^{(k)})\left[-\frac{w_{i}D_{\zeta}(y_{i};\widehat{\mu}(\bs x_{i}))}{2\phi(\bs x_{i})}-\frac{1}{2}\log(\phi(\bs x_{i}))\right],\label{eq:em-phi-update}
\end{equation}
which can be viewed as a weighted gamma regression with mean function
$\phi(\bs x)$ and shape parameter $1/2$, where the response values
are $w_{i}D_{\zeta}(y_{i};\widehat{\mu}(\bs x_{i}))$ and the observational
weights are $(1-\Pi_{i}^{(k)}),i=1,\ldots,n$.

We summarize the EM algorithm for training the zero-inflated Tweedie
model in Algorithm \ref{algo:em-fix-p}. Note that since the E and
M steps are already iterated, we no longer alternately update $F_{\mu}$
and $F_{\phi}$ in the M step, which makes the implementation easier.
The algorithm hence can be viewed as a generalized EM algorithm \citep{borman2004expectation}.
We discuss more implementation details in the sequel.
\begin{center}
\begin{algorithm}
\caption{Generalized EM algorithm for training the zero-inflated Tweedie model.}
\label{algo:em-fix-p}
\DontPrintSemicolon
\SetKwInOut{Input}{Input}\SetKwInOut{Output}{Output}
\SetKwInOut{Initial}{Initialization}\SetKwInOut{Return}{Return}

\Input{Training data $(y_{i},\bs{x}_{i},w_{i})_{i=1}^{n}$, power parameter $\zeta$}
\Output{Estimates $\widehat{\theta}=(\widehat{F}_{\mu},\widehat{F}_{\phi},\widehat{F}_{\pi})$}
\Initial{$\widehat{\theta}^{(0)}=(\widehat{F}_{\mu}^{(0)},\widehat{F}_{\phi}^{(0)},\widehat{F}_{\pi}^{(0)})$}
\For{$k=0,1,2,\ldots,K$}{
  \textbf{E step.} Calculate the posterior mean of the latent variable
  \[
     \Pi_{i}^{(k)}=
     \frac{\widehat{\pi}^{(k)}(\bs x_{i})}{\widehat{\pi}^{(k)}(\bs x_{i})+[1-\widehat{\pi}^{(k)}(\bs x_{i})]\exp\left(-\frac{w_{i}[\widehat{\mu}^{(k)}(\bs x_{i})]^{2-\zeta}}{\widehat{\phi}^{(k)}(\bs x_{i})(2-\zeta)}\right)}
     I(y_{i}=0),i=1,\ldots,n.
  \]
  \;
  \textbf{M step.} Update the function parameters $\theta$ by
  $\widehat{\theta}^{(k+1)}=(\widehat{F}_{\mu}^{(k+1)},\widehat{F}_{\phi}^{(k+1)},\widehat{F}_{\pi}^{(k+1)})$ via gradient boosted trees, where
  \[
     \begin{split}
       \widehat{F}_{\pi}^{(k+1)}&=\argmax_{F_{\pi}}
       \sum_{i=1}^{n}\left[\Pi_{i}^{(k)}\log(\pi(\bs x_{i}))+(1-\Pi_{i}^{(k)})\log(1-\pi(\bs x_{i}))\right],\\
       \widehat{F}_{\mu}^{(k+1)}&=\argmax_{F_{\mu}}
       -\frac{1}{2}\sum_{i=1}^{n}\frac{(1-\Pi_{i}^{(k)})w_{i}}{\widehat{\phi}^{(k)}(\bs x_{i})}D_{\zeta}(y_{i};\mu(\bs x_{i})),\\
       \widehat{F}_{\phi}^{(k+1)}&=\argmax_{F_{\phi}}
       \sum_{i=1}^{n}(1-\Pi_{i}^{(k)})\left[-\frac{w_{i}D_{\zeta}(y_{i};\widehat{\mu}^{(k+1)}(\bs x_{i}))}{2\phi(\bs x_{i})}-\frac{1}{2}\log(\phi(\bs x_{i}))\right].
     \end{split}
  \]
}
\Return{$\widehat{\theta}=(\widehat{F}_{\mu}^{(K)},\widehat{F}_{\phi}^{(K)},\widehat{F}_{\pi}^{(K)})$}
\end{algorithm}
\par\end{center}

\subsection{Initialization\protect\label{subsec:em-Initialization}}

The EM algorithm has the reputation of being slow in many applications.
Therefore, it is important to use sensible initialization for our
zero-inflated Tweedie model. To that end, for each fixed $\zeta\in(1,2)$,
we first compute
\[
\mu_{0}=\frac{\sum_{i\colon y_{i}>0}w_{i}y_{i}}{\sum_{i\colon y_{i}>0}w_{i}}\enskip\text{and}\enskip\phi_{0}=\frac{\sum_{i\colon y_{i}>0}w_{i}D_{\zeta}(y_{i};\mu_{0})}{\sum_{i=1}^{n}I(y_{i}>0)}.
\]
Note from (\ref{eq:zit-model}) that $(Y_{i}|Y_{i}>0)$ follows the
positive (zero-truncated) Tweedie distribution with density form
\[
d_{\text{PTD}}(y;\mu_{i},\phi_{i}/w_{i},\zeta)=\frac{\exp\left\{ \frac{w_{i}}{\phi_{i}}\left(\frac{y\mu_{i}^{1-\zeta}}{1-\zeta}-\frac{\mu_{i}^{2-\zeta}}{2-\zeta}\right)+c\left(y,\frac{\phi_{i}}{w_{i}},\zeta\right)\right\} }{1-\exp\left\{ -\frac{w_{i}\mu_{i}^{2-\zeta}}{\phi_{i}(2-\zeta)}\right\} }I(y>0).
\]
Using $\phi_{0}$ as the initial value for the $\phi_{i}$'s, we initialize
$F_{\mu}(\cdot)$ by fitting the positive Tweedie model
\begin{equation}
\begin{aligned}\widehat{F}_{\mu}^{(0)}=\argmax_{F_{\mu}}\sum_{i\colon y_{i}>0}\biggl\{ & \frac{w_{i}}{\phi_{0}}\left(\frac{y_{i}[\exp(F_{\mu}(\bs x_{i}))]^{1-\zeta}}{1-\zeta}-\frac{[\exp(F_{\mu}(\bs x_{i}))]^{2-\zeta}}{2-\zeta}\right)\\
 & -\log\left[1-\exp\left(-\frac{w_{i}[\exp(F_{\mu}(\bs x_{i}))]^{2-\zeta}}{\phi_{0}(2-\zeta)}\right)\right]\biggr\}
\end{aligned}
\label{eq:pos-tweedie}
\end{equation}
and compute $\widehat{\phi}_{0}=\sum_{i\colon y_{i}>0}w_{i}D_{\zeta}(y_{i};\exp(\widehat{F}_{\mu}^{(0)}(\bs x_{i})))/\sum_{i=1}^{n}I(y_{i}>0)$
and 
\[
\widehat{\pi}_{0}=\frac{\sum_{i=1}^{n}I(y_{i}=0)-\sum_{i=1}^{n}\exp\left(-\frac{w_{i}[\exp(\widehat{F}_{\mu}^{(0)}(\bs x_{i}))]^{2-\zeta}}{\widehat{\phi}_{0}(2-\zeta)}\right)}{n-\sum_{i=1}^{n}\exp\left(-\frac{w_{i}[\exp(\widehat{F}_{\mu}^{(0)}(\bs x_{i}))]^{2-\zeta}}{\widehat{\phi}_{0}(2-\zeta)}\right)}.
\]
If the fraction of zeros is smaller than the fitted Tweedie model
predicts, there is no reason to fit a zero-inflated Tweedie model.
We then set $\widehat{F}_{\pi}^{(0)}(\bs x)\equiv\text{logit}(\widehat{\pi}_{0})$
and $\widehat{F}_{\phi}^{(0)}(\bs x)\equiv\log(\widehat{\phi}_{0})$.
Empirically, we have seen that $\widehat{F}_{\mu}^{(0)}$ serves as
an excellent guess for $\widehat{F}_{\mu}$. We describe how to find
$\widehat{F}_{\mu}^{(0)}$ using gradient boosting in Section \ref{subsec:gbdt}.

\subsection{Function estimation via gradient boosted trees\protect\label{subsec:gbdt}}

Gradient boosting is a powerful tool for nonparametric estimation.
Decision-tree-based gradient boosting inherits many advantages from
the decision trees while having very competitive predictive performance
due to ensembles and sequential learning \citep{friedman01_greedy_functi_approx_gradie_boosti_machin,hastie2009_elemen_statis_learni,schapire14_boosting_founda_algori}.
For example, gradient boosted trees can easily handle numeric and
categorical variables without too much pre-processing, and are robust
to outliers and missing values in these variables. In this section,
we briefly describe the general decision-tree-based gradient boosting
and discuss in detail how the updates in (\ref{eq:em-pi-update}),
(\ref{eq:em-mu-update}), (\ref{eq:em-phi-update}) and (\ref{eq:pos-tweedie})
can be obtained by gradient boosting.

Consider function estimation under a general loss function $L(y,F(\bs x))$,
where the target of estimation is
\[
F^{*}=\argmin_{F}\ev_{Y,\bs X}[L(Y,F(\bs X))]=\argmin_{F}\ev_{\bs X}[\ev_{Y|\bs X}\{L(Y,F(\bs X))|\bs X\}].
\]
 Given independent observations $(y_{i},\bs x_{i}),i=1,\ldots,n$,
the boosting method fits an additive model to estimate $F^{*}$ by
\[
\min_{\{\beta_{m},\bs{\xi}_{m}\}_{m=1}^{M}}\sum_{i=1}^{n}L(y_{i},F(\bs x_{i})),\enskip F(\bs x)=\sum_{m=1}^{M}\beta_{m}B(\bs x;\bs{\xi}_{m}),
\]
where the base learners $B(\bs x;\bs{\xi}_{m})$ are simple functions
of the covariates $\bs x$ with parameters $\bs{\xi}_{m},m=1,\ldots,M$.
In practice, we often take $B(\bs x;\bs{\xi}_{m})$ to be decision
trees
\[
B(\bs x;\bs{\xi}_{m})=\sum_{j=1}^{J_{m}}\gamma_{jm}I(\bs x\in R_{jm}),m=1,\ldots,M,
\]
where the $R_{jm}$'s denote the terminal regions of the covariate
space split according to the decision tree, and the $\gamma_{jm}$'s
are constant parameters. Following \citet{friedman01_greedy_functi_approx_gradie_boosti_machin},
$F(\bs x)$ is learned in a greedy forward stagewise manner, i.e.,
in the $m$th step, we fix $F_{m-1}$ from the previous step and compute
\begin{equation}
(\beta_{m},\bs{\xi}_{m})=\argmin_{\beta,\bs{\xi}}\sum_{i=1}^{n}L(y_{i},F_{m-1}(\bs x_{i})+\beta B(\bs x_{i};\bs{\xi})),\label{eq:fw-stagewise}
\end{equation}
and then update $F$ by $F_{m}(\bs x)=F_{m-1}(\bs x)+\beta_{m}B(\bs x;\bs{\xi}_{m})$
for $m=1,\ldots,M$. In practice, (\ref{eq:fw-stagewise}) is approximately
solved by second order expansion which corresponds to a gradient descent
update. When $B(\bs x;\bs{\xi})$ is a decision tree, we often adopt
the following general scheme of gradient boosted trees for learning
$F^{*}$ \citep[see, e.g.,][]{hastie2009_elemen_statis_learni}.
\begin{enumerate}
\item Initialize $F$ with a constant
\[
F_{0}(\bs x)\equiv\argmin_{\gamma}\sum_{i=1}^{n}L(y_{i},\gamma).
\]
\item For $m=1$ to $M$:
\begin{enumerate}
\item For $i=1,\ldots,n$, compute
\[
g_{im}=\left.-\frac{\partial L(y_{i},F(\bs x_{i}))}{\partial F}\right|_{F=F_{m-1}}.
\]
\item Fit a regression tree to the targets $g_{im}$'s giving terminal regions
$R_{jm},j=1,\ldots,J_{m}$.
\item For $j=1,\ldots,J_{m}$, compute 
\begin{equation}
\widehat{\gamma}_{jm}=\argmin_{\gamma}\sum_{i\colon\bs x_{i}\in R_{jm}}L(y_{i},F_{m-1}(\bs x_{i})+\gamma).\label{eq:gbdt-terminal-est}
\end{equation}
\item Update $F_{m}(\bs x)=F_{m-1}(\bs x)+\nu\cdot\sum_{j=1}^{J_{m}}\widehat{\gamma}_{jm}I(\bs x\in R_{jm})$
for some $\nu\in(0,1)$.
\end{enumerate}
\item Output $\widehat{F}(\bs x)=F_{m}(\bs x)$.
\end{enumerate}
The constant fit $\widehat{\gamma}_{jm}$ in each terminal region
may be obtained approximately, e.g., by a second order approximation
of the loss function when it is complicated. In practice, the learning
rate $\nu$ and the base tree size $J_{m}$ are often chosen small
to avoid overfitting \citep{friedman01_greedy_functi_approx_gradie_boosti_machin,hastie2009_elemen_statis_learni}. 

Now to estimate $F_{\pi}$ from (\ref{eq:em-pi-update}), note that
$\pi(\bs x)=[1+\exp(F_{\pi}(\bs x))]^{-1}$. Therefore, the corresponding
loss function is
\[
L_{\pi}(y,F(\bs x))=-y\log\left[\frac{\exp(F(\bs x))}{1+\exp(F(\bs x))}\right]-(1-y)\log\left[\frac{1}{1+\exp(F(\bs x))}\right],
\]
which is known as the cross-entropy loss, where the target $y$ can
be anywhere between $0$ and $1$. It can be seen that the functional
gradient of $L_{\pi}$ is
\[
\frac{\partial L_{\pi}(y,F(\bs x))}{\partial F}=\frac{\exp(F(\bs x))}{1+\exp(F(\bs x))}-y.
\]
To estimate $F_{\mu}$ from (\ref{eq:em-mu-update}), note that $\mu(\bs x)=\exp(F_{\pi}(\bs x))$
and hence the corresponding loss function is
\[
L_{\mu}(y,F(\bs x))=\omega D_{\zeta}(y,\exp(F(\bs x))),
\]
whose functional gradient is
\[
\frac{\partial L_{\mu}(y,F(\bs x))}{\partial F}=2\omega\left[\exp((2-\zeta)F(\bs x))-y\exp((1-\zeta)F(\bs x))\right],
\]
where $\omega\geq0$ denotes the generic observational weight. To
estimate $F_{\phi}$ from (\ref{eq:em-phi-update}), note that $\phi(\bs x)=\exp(F_{\phi}(\bs x))$
and thus the corresponding loss function can be written as
\[
L_{\phi}(y,F(\bs x))=\omega\left[\frac{y}{\exp(F(\bs x))}+F(\bs x)\right],
\]
whose functional gradient is
\[
\frac{\partial L_{\phi}(y,F(\bs x))}{\partial F}=\omega\left[1-y\exp(-F(\bs x))\right],
\]
where $\omega$ is the observational weight. Finally, to initialize
$F_{\mu}$ from the positive Tweedie model (\ref{eq:pos-tweedie}),
we employ the loss function
\begin{align*}
L_{0}(y,F(\bs x))=\omega\biggl( & \frac{y[\exp(F(\bs x))]^{1-\zeta}}{1-\zeta}-\frac{[\exp(F(\bs x))]^{2-\zeta}}{2-\zeta}\biggr)\\
 & -\log\left[1-\exp\left(-\frac{\omega[\exp(F(\bs x))]^{2-\zeta}}{2-\zeta}\right)\right],
\end{align*}
whose functional gradient is
\[
\frac{\partial L_{0}(y,F(\bs x))}{\partial F}=w\exp[(1-\zeta)F(\bs x)]\left\{ \frac{\exp(F(\bs x))}{1-\exp\left(-\frac{\omega\exp[(2-\zeta)F(\bs x)]}{2-\zeta}\right)}-y\right\} ,
\]
where $\omega=w/\phi_{0}$ with $w$ being the exposure.

For the actual implementation, we rely on the general gradient boosting
framework of LightGBM \citep{ke2017lightgbm}, which is highly efficient
and capable of handling large-scale data. When available, LightGBM
can also use the Hessian information of the loss function to accelerate
the computation. For example, for $L_{\pi},L_{\mu}$ and $L_{\phi}$,
we have
\[
\frac{\partial^{2}L_{\pi}(y,F(\bs x))}{\partial F^{2}}=\frac{\exp(F(\bs x))}{[1+\exp(F(\bs x))]^{2}},
\]
\[
\frac{\partial^{2}L_{\mu}(y,F(\bs x))}{\partial F^{2}}=2\omega\left[(2-\zeta)\exp((2-\zeta)F(\bs x))-(1-\zeta)y\exp((1-\zeta)F(\bs x))\right],
\]
and
\[
\frac{\partial^{2}L_{\phi}(y,F(\bs x))}{\partial F^{2}}=\omega y\exp(-F(\bs x)).
\]

\subsection{Power parameter estimation}

The power parameter of the Tweedie model is often estimated by the
profile likelihood \citep{dunn05_series_evaluat_tweed_expon_disper_model_densit}.
For the zero-inflated Tweedie model, we can similarly employ the profile
likelihood to estimate the power parameter $\zeta$ of the Tweedie
state. Specifically, we choose a reasonable sequence of possible values
$\{\zeta_{1},\ldots,\zeta_{K}\}$ for $\zeta$. For each $\zeta_{k},k=1,\ldots,K$,
we apply the EM algorithm from Algorithm (\ref{algo:em-fix-p}) to
estimate the function parameters $\widehat{\theta}_{k}=(\widehat{F}_{\mu,k},\widehat{F}_{\phi,k},\widehat{F}_{\pi,k})$.
We then calculate the attained log-likelihood $\ell(\widehat{\theta}_{k},\zeta_{k}|\bs y,\bs X,\bs w)$
from (\ref{eq:zit-loglik}) using the training data. The best power
parameter $\widehat{\zeta}$ is then obtained as $\zeta_{\widehat{k}}$,
where
\[
\widehat{k}=\argmax_{1\leq k\leq K}\ell(\widehat{\theta}_{k},\zeta_{k}|\bs y,\bs X,\bs w).
\]
We summarize the profile likelihood estimation of $\zeta$ in Algorithm
\ref{algo:profile-lik-power}.
\begin{center}
\begin{algorithm}
\caption{Profile likelihood estimation in the zero-inflated Tweedie model.}
\label{algo:profile-lik-power}
\DontPrintSemicolon
\SetKwInOut{Input}{Input}\SetKwInOut{Output}{Output}
\SetKwInOut{Initial}{Initialization}\SetKwInOut{Return}{Return}

\Input{Training data $(y_{i},\bs{x}_{i},w_{i})_{i=1}^{n}$}
\Output{Estimates $\widehat{\zeta}$ and $\widehat{\theta}=(\widehat{F}_{\mu},\widehat{F}_{\phi},\widehat{F}_{\pi})$}
\Initial{Choose a sequence of candidate power parameters $\{\zeta_{1},\ldots,\zeta_{K}\}\subset(1,2)$}
\For{$k=1,2,\ldots,K$}{
   Set $\zeta=\zeta_{k}$.\;
   Apply Algorithm~\ref{algo:em-fix-p} to obtain estimates $\widehat{\theta}_{k}=(\widehat{F}_{\mu,k},\widehat{F}_{\phi,k},\widehat{F}_{\pi,k})$.\;
   Calculate the log-likelihood $\ell(\widehat{\theta}_{k},\zeta_{k}|\bs y,\bs X,\bs w)$.\;
}
\Return{$\widehat{\zeta}=\zeta_{\widehat{k}}$ and $\widehat{\theta}=\widehat{\theta}_{\widehat{k}}$, where
   \[
      \widehat{k}=\argmax_{1\leq k\leq K}\ell(\widehat{\theta}_{k},\zeta_{k}|\bs y,\bs X,\bs w).
   \]
}
\end{algorithm}
\par\end{center}

\section{Numerical studies\protect\label{sec:numerical-studies}}

In this section, we conduct simulation experiments and real data analysis
to demonstrate the performance of our proposed approach, which we
coin as ZITboost. For comparison, we consider the Tweedie boosting
method \citep[TDboost,][]{yang2018insurance} and the zero-inflated
Tweedie boosting method \citep[EMTboost,][]{zhou20_tweed_gradien_boost_extrem_unbal}.

\subsection{Simulation experiments}

We simulate data from a zero-inflated Tweedie model based on the random
function generator of \citet{friedman01_greedy_functi_approx_gradie_boosti_machin}.
Specifically, let the covariates $\bs x\sim N(\bs 0,\bs I_{p})$,
where $\bs I_{p}$ is the identity matrix of size $p\times p$ and
we set the dimension $p=10$. The random function generator is often
used to generate highly random target functions in model specification
and can help avoid certain simulation setups that may favor one method
over another. It takes the form
\begin{equation}
F^{*}(\bs x)=\sum_{j=1}^{20}a_{j}g_{j}(\bs z_{j}),\label{eq:rfg}
\end{equation}
where each coefficient of $(a_{j})_{j=1}^{20}$ is independently drawn
from a uniform distribution $U(-1,1)$, and each $g_{j}(\bs z_{j})$
is a function of a randomly selected subset, of size $p_{j}$, of
the $p$ covariates $\bs x$, that is, $\bs z_{j}=(x_{\sigma_{j}(k)})_{k=1}^{p_{j}}$,
where $\sigma_{j}$ denotes an independent random permutation of the
indices $\{1,\ldots,p\}$. The subset size $p_{j}=\min(\lfloor1.5+r_{j}\rfloor,p)$
is also randomly generated, where $r_{j}$ is independently drawn
from an exponential distribution with mean $\lambda=5$. Then, $g_{j}(\bs z_{j})$
is constructed from the $p_{j}$-dimensional Gaussian function
\[
g_{j}(\bs z_{j})=\exp\left(-\frac{1}{2}(\bs z_{j}-\bs{\mu}_{j})^{\T}\bs V_{j}(\bs z_{j}-\bs{\mu}_{j})\right),
\]
where $\bs{\mu}_{j}$ is drawn from $N(\bs 0,\bs I_{p_{j}})$ and
the precision matrix $\bs V_{j}=\bs U_{j}\bs D_{j}\bs U_{j}^{\T}$
is also randomly generated, with $\bs U_{j}$ being a random orthogonal
matrix and $\bs D_{j}=\text{diag}(d_{1j},\ldots,d_{p_{j}j})$ for
independent $\sqrt{d_{kj}}\sim U(0.1,2),k=1,\ldots,p_{j}$.

\subsection{Real data analysis}

We consider the \textbf{AutoClaim} data from the \texttt{cplm} R package.
This motor insurance data set was retrieved from the SAS Enterprise
Miner database. It contains 10296 records and 28 variables. The data
were previously studied in \citet{yip2005modeling} and \citet{zhang12_likel_based_bayes_method_tweed}.
See also \citet{yang2018insurance} for a description of the variables
available in the data. Our objective is to estimate the expected pure
premium by leveraging the predictor variables. Around 61.1\% of the
individuals insured had no claims, while 29.6\% of policyholders had
positive claim amounts not exceeding \$10,000. A small fraction, specifically
9.3\%, of the policyholders, experienced high claim amounts surpassing
\$10,000. The cumulative sum of the high claim amounts accounted for
approximately 64\% of the total sum. Following \citet{zhou20_tweed_gradien_boost_extrem_unbal},
we employ this original data set to generate synthesized scenarios
that reflect the often encountered and more realistic cases characterized
by excessively zero-inflated data sets. Given that the initial proportion
of zero claim amounts (61.1\%) is not sufficiently extreme, we implement
a sampling approach to enhance the predominance of zero claims. To
ensure comparability of results, we employ a similar data preprocessing
method as described in the study by \citet{zhou20_tweed_gradien_boost_extrem_unbal}.
We perform random under-sampling (without replacement) from the subset
of nonzero-claim data using a fraction of 0.15. This procedure effectively
augments the percentage of zero-claim policyholders, resulting in
an approximate increase to 91.28\% of the total population. After
performing the aforementioned sampling procedure, we proceed to partition
the resulting dataset into two distinct sets; the training and testing
sets, using a uniform allocation approach. The training dataset comprises
3531 observations, with approximately 91.62\% of the claim amounts
being zero. Similarly, the testing dataset consists of 3360 observations,
wherein approximately 90.92\% of the claim amounts are zero.

To evaluate and compare the predictive performance of TDboost, EMTboost,
and our method, we employ them to predict the pure premium on the
held-out testing dataset. Various performance metrics, including mean
squared error (MSE), mean absolute deviation (MAD), and deviance,
are utilized to assess the accuracy of the predictions. Given the
highly right-skewed nature of the losses, we utilize the ordered Lorentz
curve and the associated Gini index to effectively capture the disparities
between the expected premiums and actual losses. These metrics provide
comprehensive insights into the model's ability to capture and quantify
the underlying discrepancies.

\section{Conclusion\protect\label{sec:conclusions}}

The zero-inflated Tweedie model provides a natural interpretation
for extremely zero-inflated insurance claim data. The dispersion and
zero-state probability modeling helps to improve the overall model
performance. The EM algorithm is straightforward to implement based
on LightGBM and the initialization from fitting the positive Tweedie
model helps reduce the number of EM iterations required.

The zero-inflated Poisson part of the zero-inflated Tweedie model
in principle can be replaced by a zero-inflated negative binomial.
In that case, one models the extremely zero-inflated aggregated claim
size by a compound negative binomial and gamma mixture. The new model
has the potential to handle scenarios where there is excess variability
in the number of claims being observed under each policy. However,
fitting this model may be computationally more difficult.

\bibliographystyle{nameyear}
\bibliography{refs}

\end{document}